\documentclass[traditabstract]{aa}
\usepackage{txfonts}
\usepackage{graphicx}
\usepackage{epsfig}
\usepackage{natbib}

\begin{document}

\title{Deep census of variable stars in a VLT/VIMOS field
in Carina\thanks{Based on observations collected with the
Very Large Telescope at Paranal Observatory (ESO Programme
075.C-0427(A), DM and JMF visiting observers).}
}
\subtitle{}
\author{P. Pietrukowicz\inst{1,2}
        \and
        D. Minniti\inst{1,3}
        \and
        J.~M. Fern\'andez\inst{1,4}
        \and
        G. Pietrzy\'nski\inst{5,6}
        \and \\
        M.~T. Ru\'iz\inst{7}
        \and
        W. Gieren\inst{5}
        \and
        R.~F. D\'iaz\inst{8}
        \and
        M. Zoccali\inst{1}
        \and
        M. Hempel\inst{1}
        }

\offprints{P. Pietrukowicz,\\ \email{pietruk@astro.puc.cl}}

\institute{Departamento de Astronom\'ia y Astrof\'isica,
Pontificia Universidad Cat\'olica de Chile, Av. Vicu\~na MacKenna 4860,
Casilla 306, Santiago 22, Chile
       \and
Nicolaus Copernicus Astronomical Center, ul. Bartycka 18, 00-716 Warszawa, Poland
       \and
Vatican Observatory, Vatican City State V-00120, Italy
       \and
Harvard-Smithsonian Center for Astrophysics, 60 Garden Street,
Cambridge, MA 02138, USA
       \and
Departamento de Astronom\'ia, Universidad de Concepci\'on, Casilla 160-C,
Concepci\'on, Chile
       \and
Warsaw University Observatory, Al. Ujazdowskie 4, 00-478 Warszawa, Poland
       \and
Departamento de Astronom\'ia, Universidad de Chile, Casilla 36-D, Santiago, Chile
       \and
Instituto de Astronom\'ia y F\'isica del Espacio (CONICET-UBA), Buenos Aires, Argentina
}

\date{Received ; accepted }

\abstract{
We have searched for variable stars in deep $V$-band images
of a field towards the Galactic plane in Carina. The images were taken
with VIMOS instrument at ESO VLT during 4 contiguous nights in April 2005.
We detected 348 variables among 50897 stars in the magnitude range between
$V=15.4$ and $V=24.5$~mag. Upon detection, we classified the variables by
direct eye inspection of their light curves. All variable objects but 9 OGLE
transits in the field are new discoveries. We provide a complete catalog
of all variables which includes eclipsing/ellipsoidal binaries,
miscellaneous pulsators (mostly $\delta$~Scuti-type variables),
stars with flares and other (irregular and likely long-period) variables.
Only two of the stars in our sample are known to host planets.
Our result give some implications for future large variability surveys.
}


\maketitle

\section{Introduction}

Variable stars provide important information on the nature and evolution
of stars in various regions of our Galaxy. Eclipsing binary stars give us
information on the masses and radii of stars \citep[see e.g. ][]{hua56,popp85}.
For some, i.e. the detached systems, it is possible to measure distances
\citep{pacz97}, which is particularly useful in case of star clusters
\citep[see e.g. ][]{bon06,kal07}. Pulsating variables, such as Cepheids
and RR~Lyrae, serve as such distance indicators \citep[e.g. ][]{ben07,fea08}.
The type of pulsators known as $\delta$~Scuti stars offer a unique
insight into the internal structure and evolution of main-sequence
objects \citep{thom03,goup05}.

Due to the advent of large-scale wide-field photometric surveys 
the number of new variables has increased rapidly.
The surveys are usually dedicated to the detection of particular objects,
like transiting extrasolar planets, microlensing events or gamma ray burst
afterglows, e.g. MACHO \citep{alco00} and OGLE surveys \citep{uda03}.
Most of the wide-field surveys are conducted with small (less than 0.5~m)
robotic telescopes, e.g. ASAS \citep[All-Sky Automated Survey,][]{poj01},
ROTSE \citep[Robotic Optical Transient Search Experiment,][]{woz04},
allowing the search for variability only for bright stars (down
to $\sim15$~mag in the $V$ band). Until recently the big telescopes, i.e.
those of 4-10 meters in diameter, have been focusing rather on individual
objects, due to their small fields of view. Practically there
have been very few long and deep variability surveys. In this paper we present
the results of searches for variable objects in a deep (down to $V\approx24.5$~mag)
galactic field in the constellation of Carina, based on data collected with an 8-m
telescope for 4 continuous nights. Our results show the great potential of
upcoming large visual and near-infrared surveys, like LSST
\citep[Large Synoptic Survey Telescope,][]{ive08}, VVV \citep[Vista
Variables in Via Lactea,][]{ahu06} or Pan-STARRS \citep{kai02}.

\section{Observations and data reduction}

Observations were carried out with VIMOS at the Unit Telescope 3 (UT3) of the
European Southern Observatory Very Large Telescope (ESO VLT) located at Paranal
Observatory from April 9 to 12, 2009. VIMOS is an imager and multi-object
spectrograph \citep{lef03}. Its field of view consists of 4~quadrants of about
$7\arcmin \times 8\arcmin$ each, separated by a cross, $2\arcmin$ wide.
The CCD size is $2048 \times 2440$ pixels with a pixel size of $0\farcs205$.

The main goal of the program was to photometrically follow-up of over 30 OGLE
transiting candidates. Some individual results have been published by \cite{fer06},
\cite{diaz07}, \cite{hoy07}, \cite{min07}, \cite{diaz08}, \cite{pie09}.
The field analyzed in this paper is one out of four VIMOS fields
monitored during the run, and is the only one which was observed during
4 full nights. Equatorial coordinates of the center of the field are
RA(2000.0)=10$^h$52$^m$56$^s$, Dec(2000.0)=-61$\degr$28$\arcmin$15$\arcsec$
or $l=289\fdg269$, $b=-1\fdg783$. The field contains 9 OGLE transits,
two of which were confirmed to be caused by planets:
OGLE-TR-111b \citep{pont04} and OGLE-TR-113b \citep{bou04,kon04}.
All 4 nights were clear throughout, with sub-arcsecond seeing during most
of the time (see Fig. 1). The dataset consists of 660 images taken only
in the $V$-band.

The periphery of each quadrant suffers from coma. Therefore, for our
analysis we cut out a slightly smaller area of $1900 \times 2100$ pixels,
covering $7\farcm18 \times 6\farcm49$. The total field in which we searched
for variable objects equals 186.3 arcmin$^2$.

The photometry was extracted with the help of the {\it Difference Image
Analysis Package} (DIAPL) written by \cite{woz00} and recently modified
by W. Pych\footnote{The package is available at
http://users.camk.edu.pl/pych/DIAPL/}. The package is an implementation
of a method developed by \cite{ala98}. To obtain better quality
of photometry we divided the field into $475 \times 525$ pixel subfields.

Reference frames were constructed by combining the 8 or the 9 best individual
images (depending on the quadrant). The profile photometry for the reference
frame was extracted with DAOPHOT/ALLSTAR \citep{stet87}. These measurements
were used to transform the light curves from differential flux units into
instrumental magnitudes, which were later transformed to the standard $V$-band
magnitudes by adding an offset derived from $V$-band magnitudes of
the planetary transits located in the field \citep{diaz07,min07}.
The quality of the photometry is illustrated in Fig.~2.
In this figure we plot the standard deviation {\it vs.} the average magnitude
for stars from one of the VIMOS quadrants.

Due to the short period of the observations we decided to look for variables initially
by direct eye inspection of all 50897 light curves. This lasted about one week.
For comparison and more reliable statistics, we performed an independent
period search with the TATRY code \citep[see][]{sch89,sch96}. After the
automatic search we sorted all stars by decreasing variability (quality)
factor, which is generated by the code, and we looked at the most promising
light curves. The total number of detected variables reached 348 objects.
Interestingly, among 175 periodic variables except $\delta$~Scuti
stars only four additional variables were found, and up to 51 variables were
missed as the result of the automatic search in comparison to the by-eye method.
The missing objects were mostly faint ($V>19$~mag) low-amplitude
($\Delta V<0.15$~mag) stars. However, some of them could be missed due to
the cut of the list of stars for revision.

Finally, all detected variables were sorted by increasing right ascension
and classified by looking at the shape of the light curves and possible
periods.

\begin{figure}
\includegraphics[angle=0,width=0.5\textwidth]{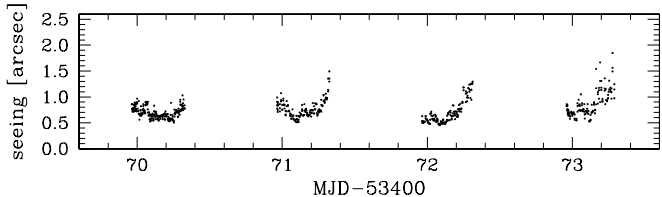}
\caption{The seeing during the four observing nights.}
\label{seeing}
\end{figure}

\begin{figure}
\includegraphics[angle=0,width=0.5\textwidth]{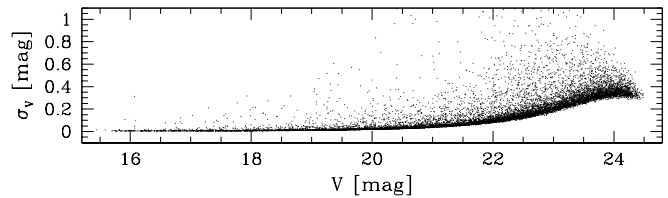}
\caption{The photometric errors for 12571 stars from the VIMOS
quadrant $A2$ plotted as a function of $V$ magnitude.}
\label{sigma}
\end{figure}

\section{The variables}

Figure~3 shows four histograms representing magnitude distributions
for all stars, all variables, eclipsing/ellipsoidal binaries and $\delta$~Scuti
stars, respectively. The distribution for the complete sample of variable
stars peaks at $V\approx18.5$~mag. The number of eclipsing/ellipsoidal
variables starts to decrease below approximately $V=19.5$~mag, but still many
faint objects of this type were detected thanks to their high amplitudes.
For $\delta$~Scuti stars, which have typical amplitudes of 0.02-0.06~mag,
the detection efficiency seems to peak at $V\approx17$~mag.
Very few stars of this type were found to be fainter than $V=20$~mag.

\begin{figure}
\includegraphics[angle=0,width=0.5\textwidth]{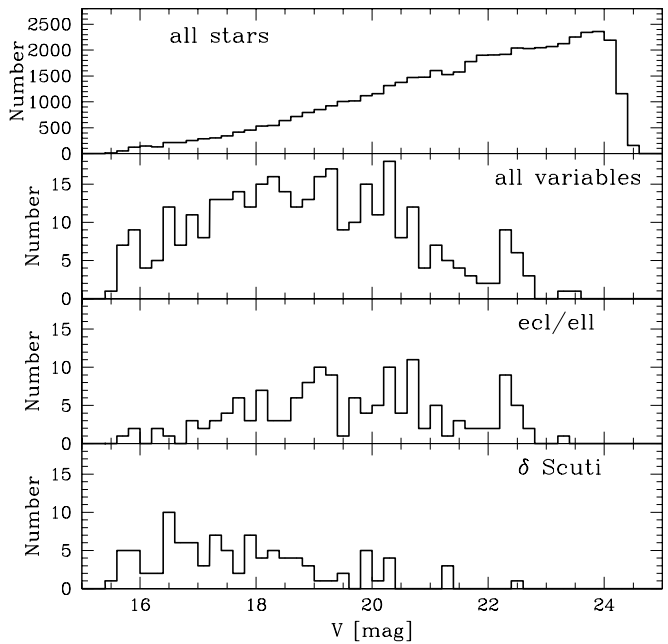}
\caption{Magnitude distribution of all analysed stars (50897 objects),
all detected variables in the sample (348 objects), all stars classified
as eclipsing/ellipsoidal binaries (148 objects) and all $\delta$~Scuti stars
(99 objects). The bin size is 0.2~mag.}
\label{magdist}
\end{figure}

\subsection{Eclipsing variables}

In Fig.~4 we show the period and amplitude distributions for eclipsing/ellipsoidal
binaries. The distribution for the shortest periods could be fit by a Gaussian with
a mean period of 0.31~d. This value differs slightly from the mean value of 0.277~d
and $\sigma=0.036$~d, as found by \cite{wel08} in a Galactic plane field in Lupus,
and the value of 0.37~d derived from ASAS data \citep{pacz06}.
From the amplitude distribution we conclude that about three fourths
of the sample binaries have $V$ amplitudes below 0.7~mag.

\begin{figure}
\includegraphics[angle=0,width=0.5\textwidth]{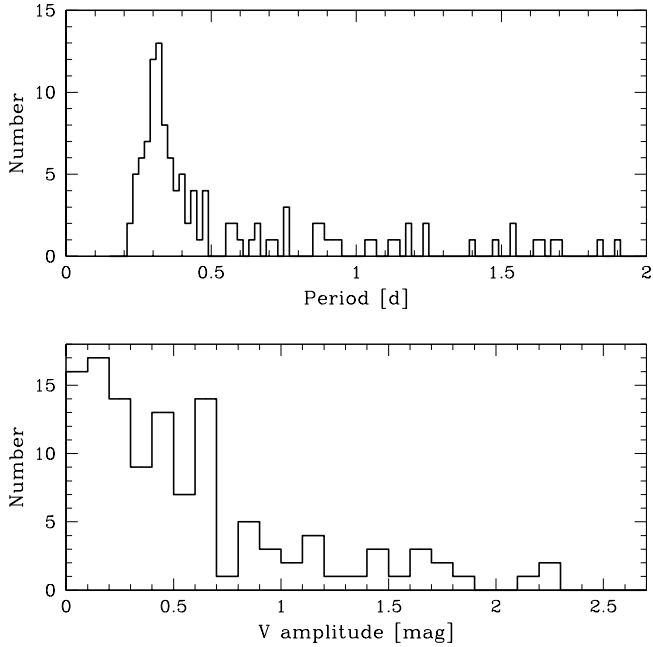}
\caption{Period distribution (upper panel) and amplitude distribution (lower
panel) for eclipsing binaries with estimated period.}
\label{perdist}
\end{figure}

The following figures, Figs.~5-7, illustrate phased light curves
for all 121 eclipsing/ellipsoidal variables, for which we were able
to derive accurate periods. Note that the faintest eclipsing variable
found in our data, V050, has the maximum brightness of about 23.3~mag
and the amplitude of $\approx1.4$~mag in $V$. System V009 has the largest
amplitude of 2.23~mag. Variable V149 shows short and shallow eclipses besides
the sinusoidal changes in its brightness. It may be an RS~CVn binary,
but more data is needed to answer the question what kind of object this is.
Variable V278 corresponds the transit OGLE-TR-109 phased with the transiting
period of 0.589127~d. The analysis of these photometric data by \cite{fer06},
combined with high-resolution spectroscopic data by \cite{pont05},
leaves the nature of the object unclear. Two scenarios are possible: OGLE-TR-109
can be either a blend of the star with a background eclipsing binary
or a transiting planet (less likely because of its very short period).

\begin{figure*}
\includegraphics[angle=0,width=1.0\textwidth]{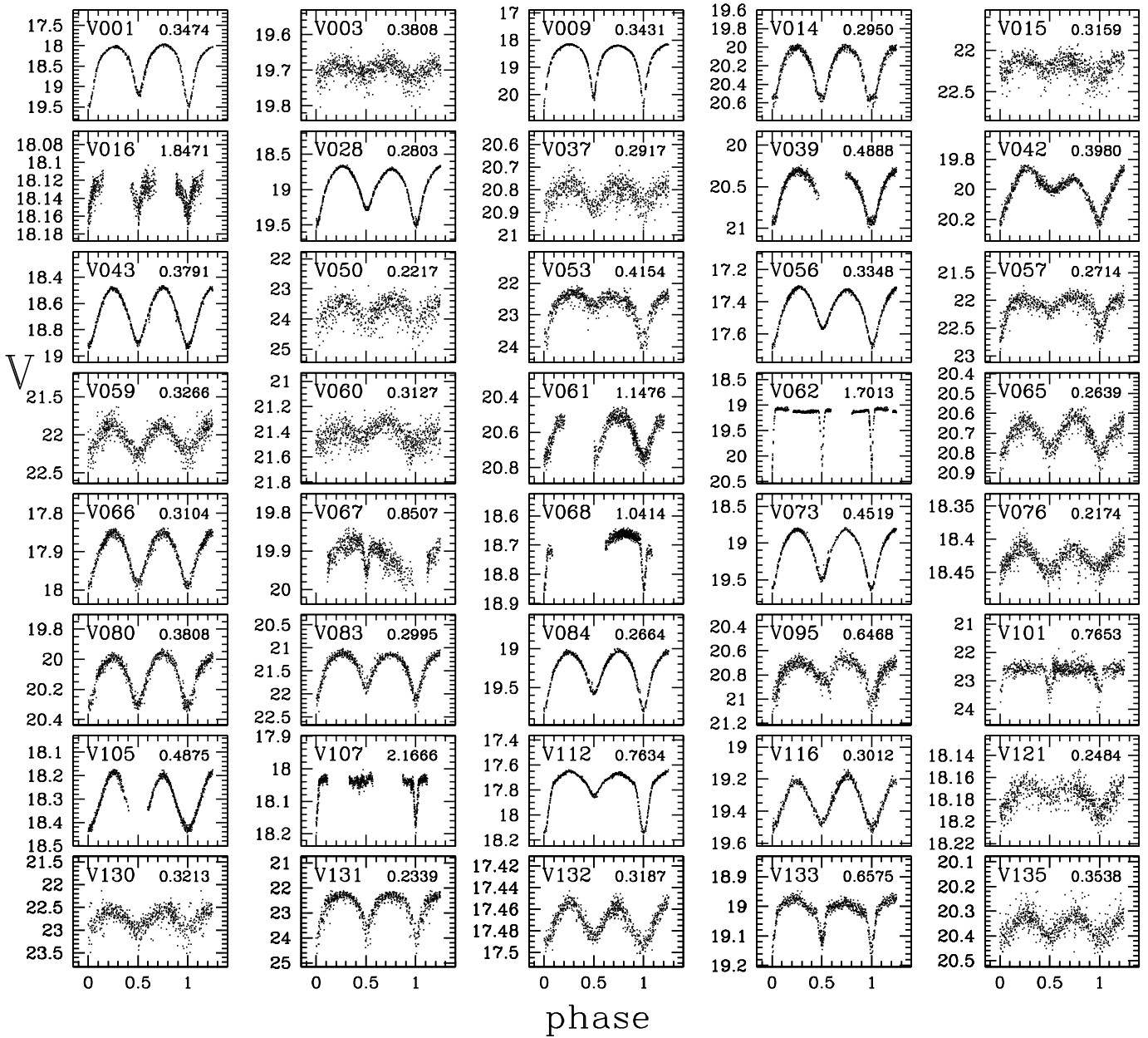}
\caption{Light curves for eclipsing or ellipsoidal variables with estimated
periods (part 1 of 3). The identifications and periods in days are given for
each object.}
\label{ecl1}
\end{figure*}

\begin{figure*}
\includegraphics[angle=0,width=1.0\textwidth]{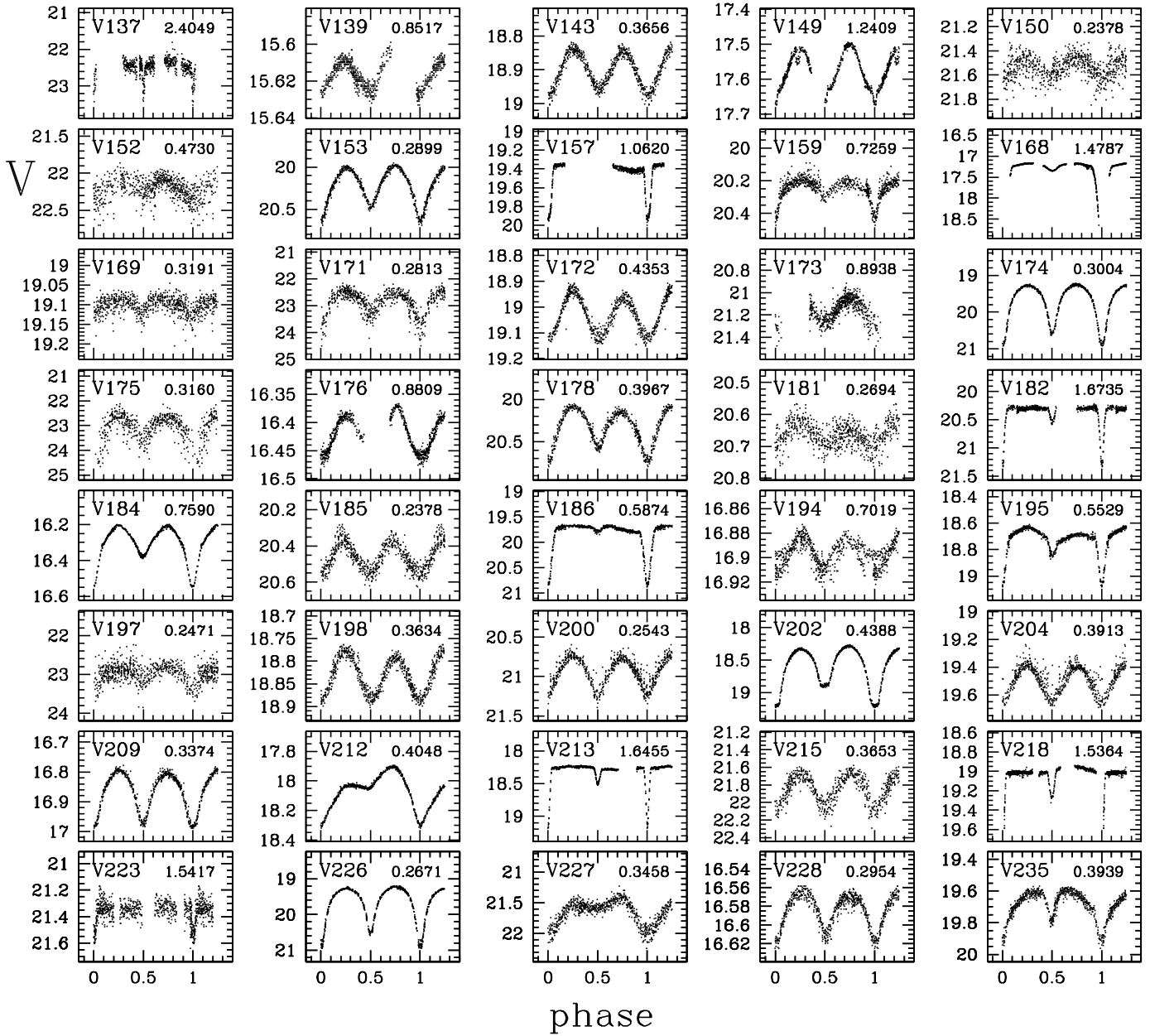}
\caption{Light curves for eclipsing or ellipsoidal variables with estimated
periods (part 2 of 3).}
\label{ecl2}
\end{figure*}

\begin{figure*}
\includegraphics[angle=0,width=1.0\textwidth]{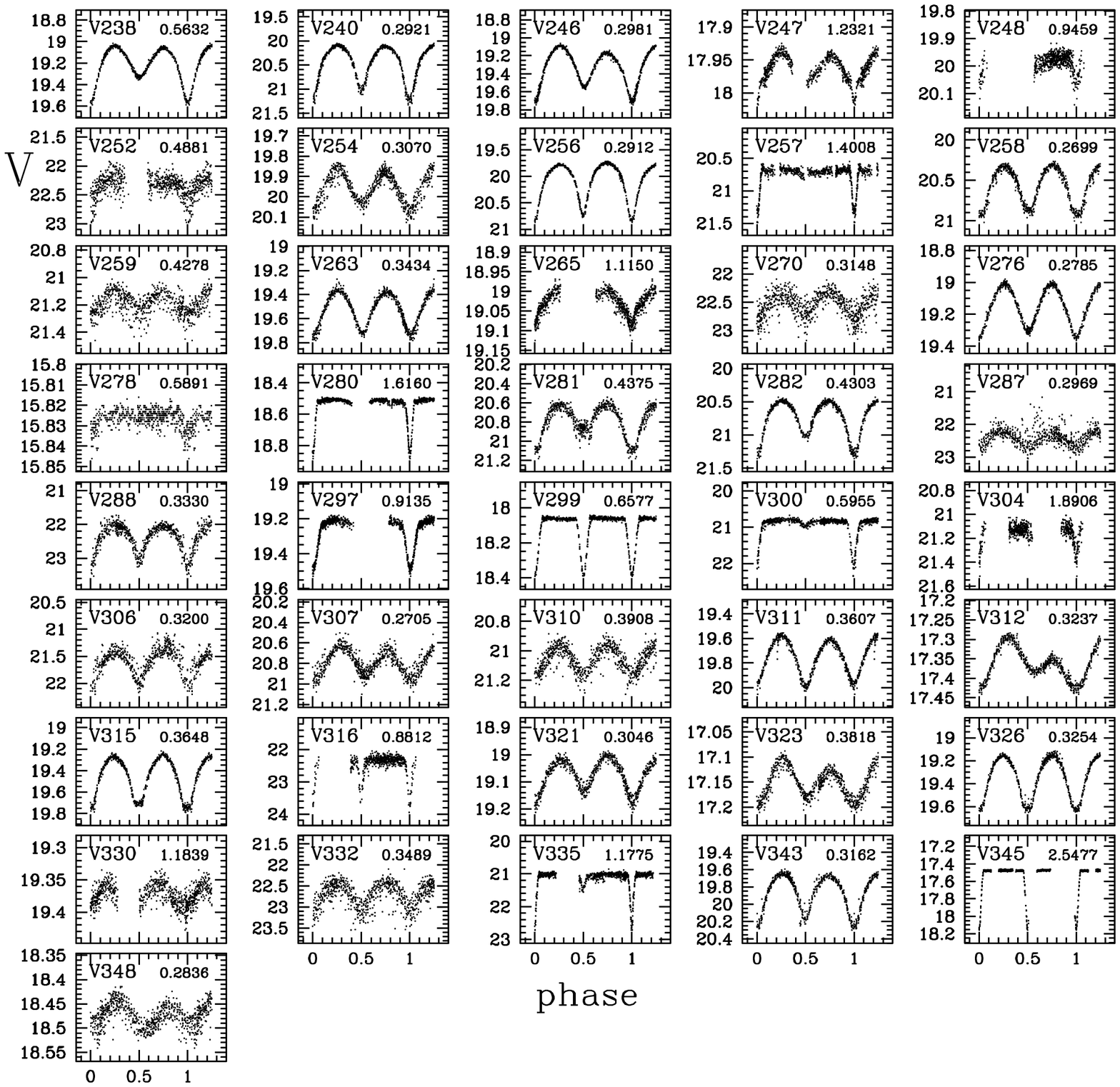}
\caption{Light curves for eclipsing or ellipsoidal variables with estimated
periods (part 3 of 3). Note that the object V278 is the OGLE transit TR-109
phased with transiting period of 0.589127~d.}
\label{ecl3}
\end{figure*}

In Fig.~8 we present the light curves of stars showing one or two eclipses,
but for which the time-span of our observations was not sufficient to estimate
the period. Some variables in this sample need further explanation. Star V314
is known as the transit OGLE-TR-106 with a period of 2.53585~days \citep{uda02}.
\cite{pont05} show that this object is not a transiting planet
but an eclipsing binary where one of the companions is a low mass
M~dwarf with $0.116 \pm 0.021$~$M_\odot$.

Object V041 changed its brightness by about 0.25~mag showing an unexpected
flat-bottom eclipse during the second night. This eclipse
lasted about 4.5~hours, had an amplitude of $\sim0.05$~mag,
and suggests a periodicity in the object. The overall shape of the
light curve indicates a periodic behavior, too.
Based on the incomplete photometric data we can only speculate about
the nature of the variable by asking several questions.
Is this object an eclipsing binary containing a pulsating star as
the primary or is it just a blend? Was the observed eclipse a transit of
a planetary/low-luminosity object or the secondary eclipse in a binary system?
If it is indeed a real binary, what are the parameters of the components?
More data are needed to find the answers in this very interesting case.

The light curve of eclipsing variable V224 features some kind of bumps
occurring regularly at the end of every night. The bumps are caused
by the diffraction spikes of a nearby saturated star.

\begin{figure*}
\includegraphics[angle=0,width=1.0\textwidth]{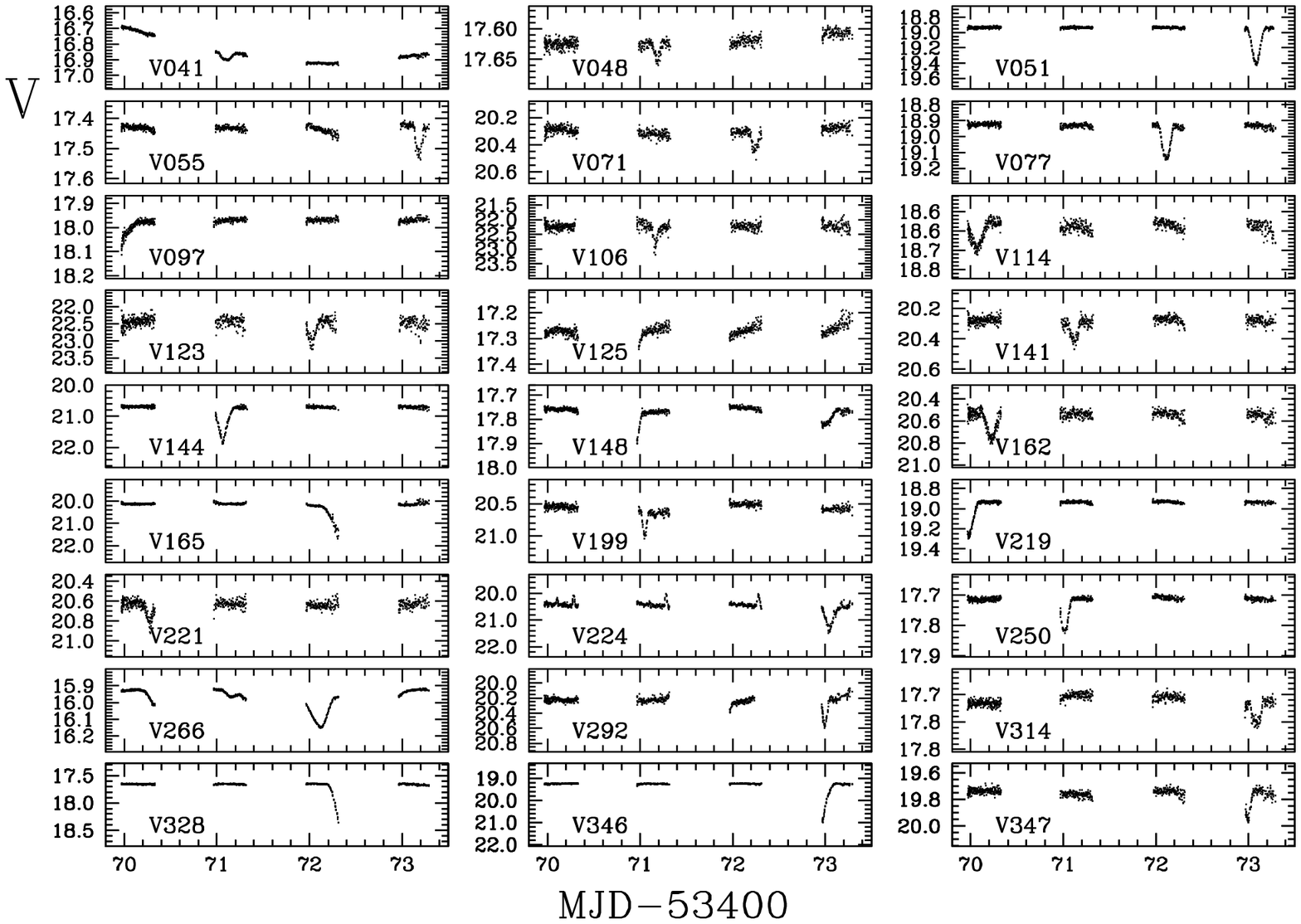}
\caption{Light curves for other eclipsing variables. Note that the object V314
is the OGLE transit TR-106 with an M dwarf star as a component.}
\label{eclother}
\end{figure*}

\subsection{Pulsating variables}

We classified 99 stars as $\delta$~Scuti type pulsators. Their light
curves are shown in Figs.~9-11. The amplitudes of all the stars except V340
are in the range between 0.015 and 0.230~mag. The variable V340
has an exceptionally large amplitude of about 1.0~mag. We do not attempt
to derive exact periods, since most of the stars are multiperiodic
variables, and hence require more sophisticated analysis, which will be
subject of future studies.

\begin{figure*}
\includegraphics[angle=0,width=1.0\textwidth]{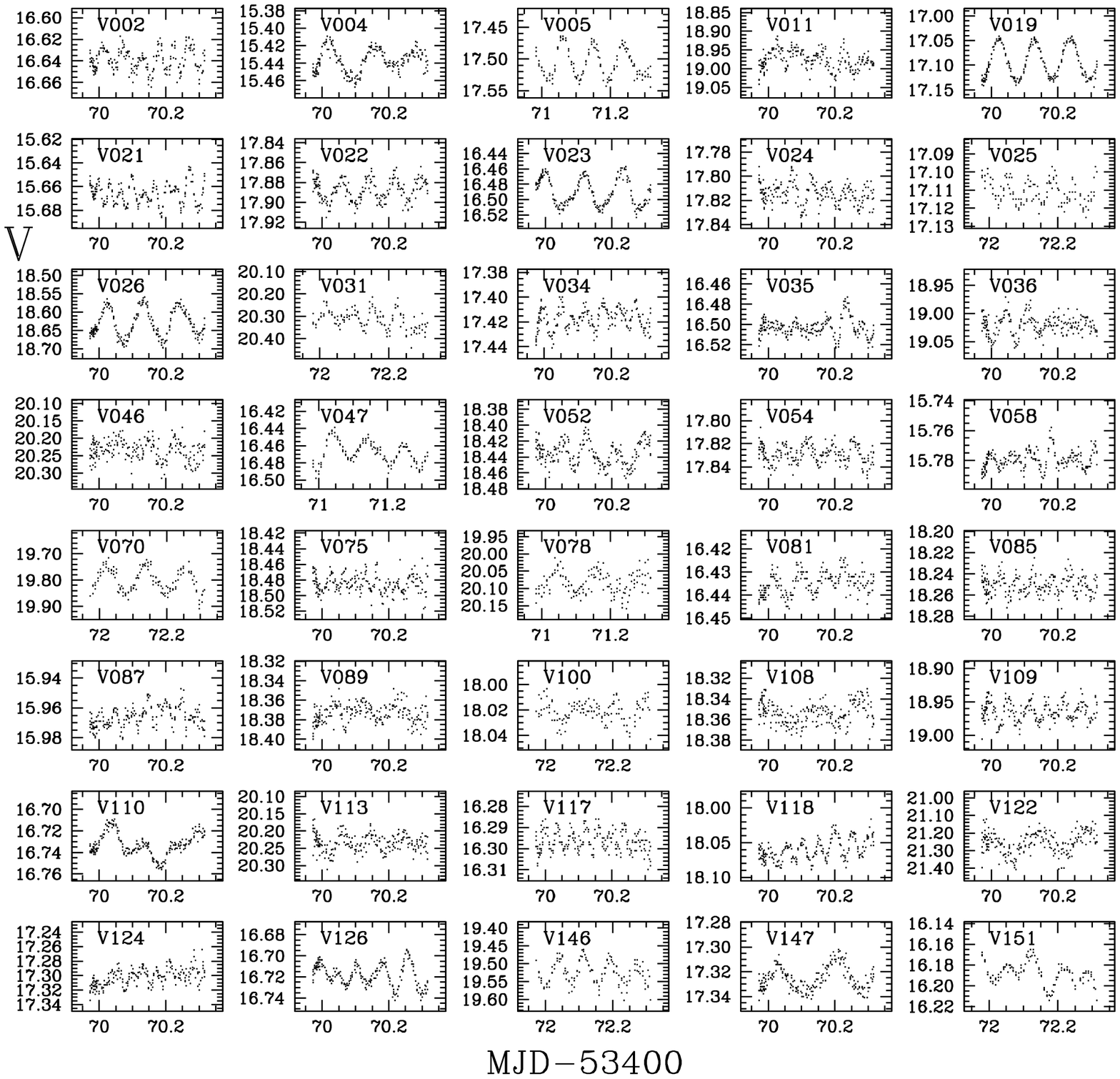}
\caption{Light curves for detected $\delta$~Scuti-type variables (part 1 of 3).
Each panel presents only data points from a single night.}
\label{delta1}
\end{figure*}

\begin{figure*}
\includegraphics[angle=0,width=1.0\textwidth]{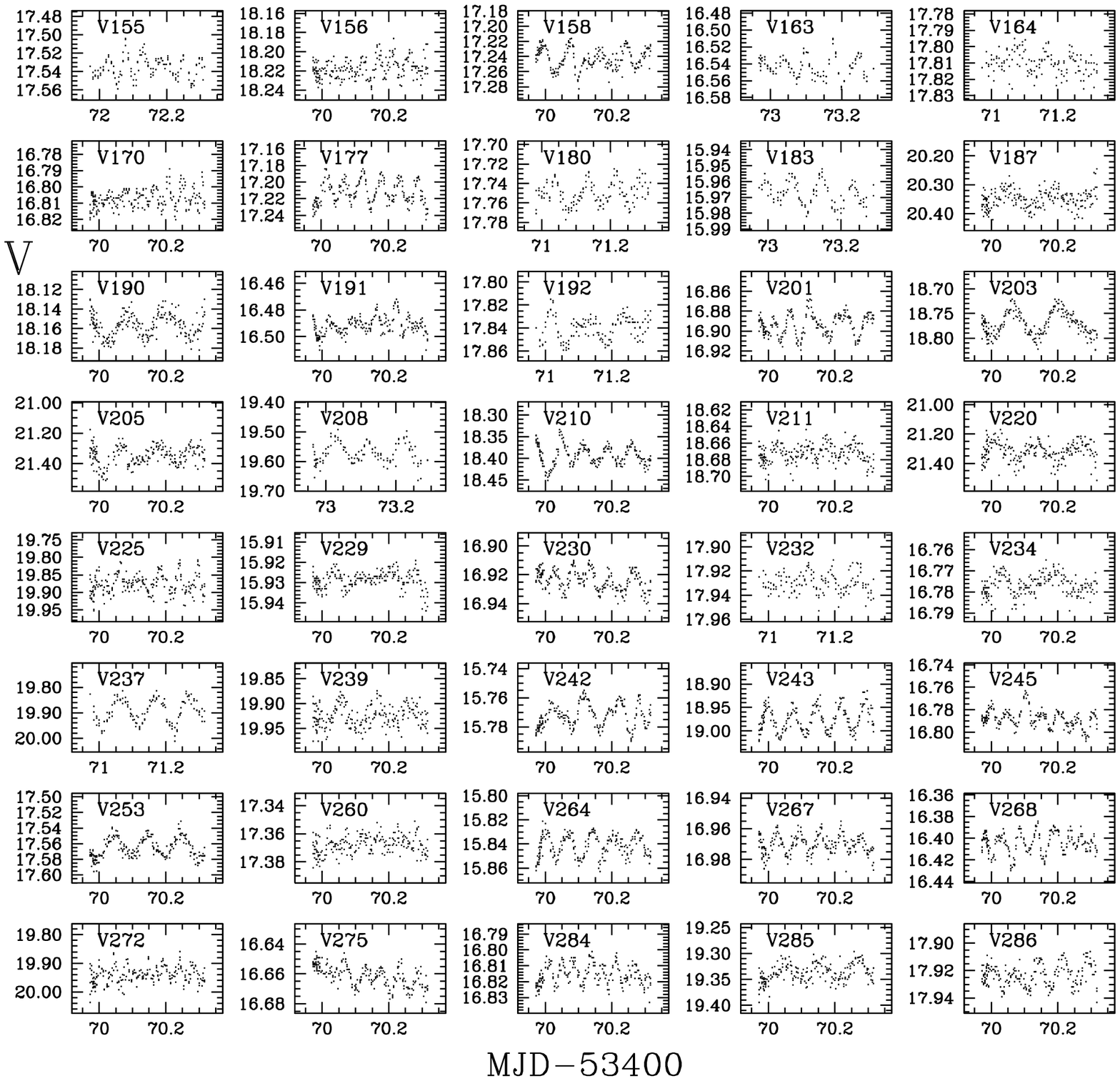}
\caption{Light curves for detected $\delta$~Scuti-type variables (part 2 of 3).}
\label{delta2}
\end{figure*}

\begin{figure*}
\includegraphics[angle=0,width=1.0\textwidth]{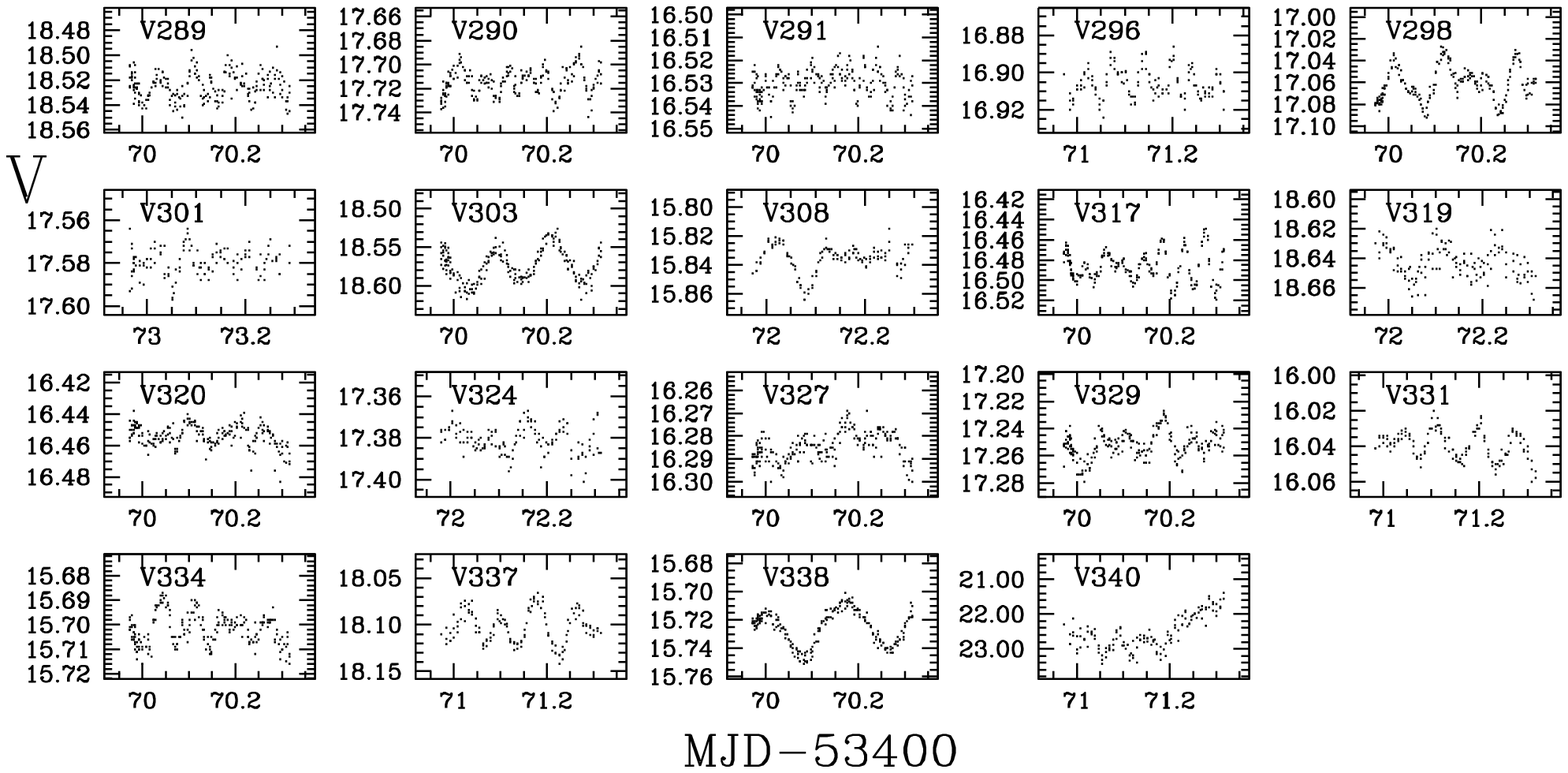}
\caption{Light curves for detected $\delta$~Scuti-type variables (part 3 of 3).}
\label{delta3}
\end{figure*}

Other pulsating variables are shown in Figs.~12-13. The periods of the
variables range between 0.2307 and 4.1527~days. Variables V010, V018,
V038, V049, V086, V098, V102, V160, V222, V251 and V342 are good candidates
for RR~Lyrae stars, whereas variables V012, V090, V231 and V339 are likely distant
Cepheids. Light curves of miscellaneous variables are shown in Fig.~14. For example,
variables V017, V029, V030, V128, V233, V241, V302 are good long-period candidates.

\begin{figure*}
\includegraphics[angle=0,width=1.0\textwidth]{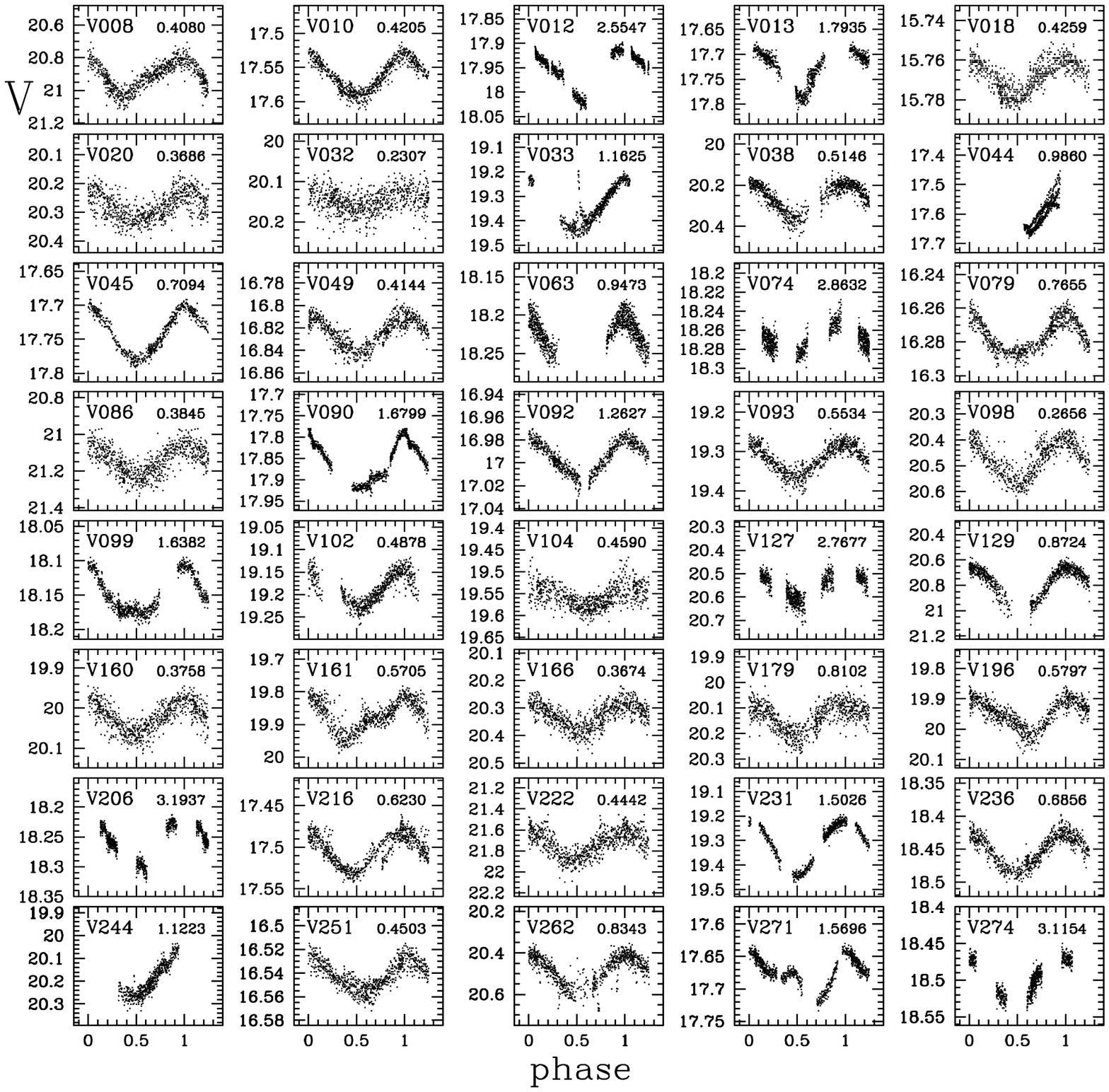}
\caption{Phased light curves for other pulsating variables with estimated periods
(part 1 of 2). The identifications and periods in days are given for each object.}
\label{puls1}
\end{figure*}

\begin{figure*}
\includegraphics[angle=0,width=1.0\textwidth]{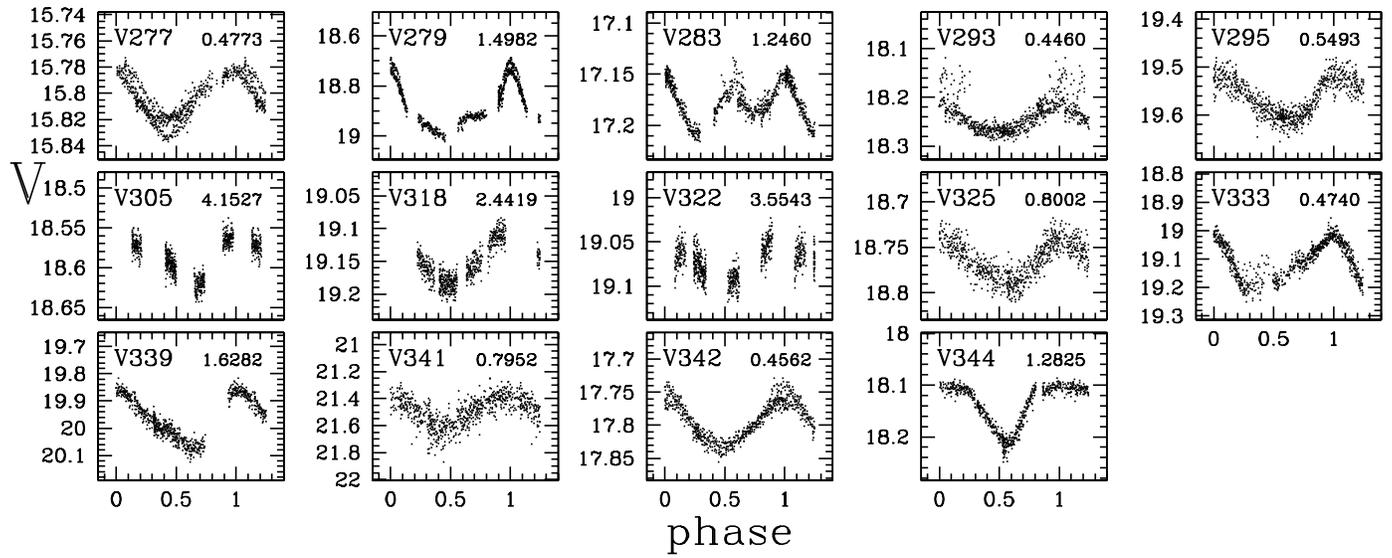}
\caption{Phased light curves for other pulsating variables with estimated periods
(part 2 of 2).}
\label{puls2}
\end{figure*}

\begin{figure*}
\includegraphics[angle=0,width=1.0\textwidth]{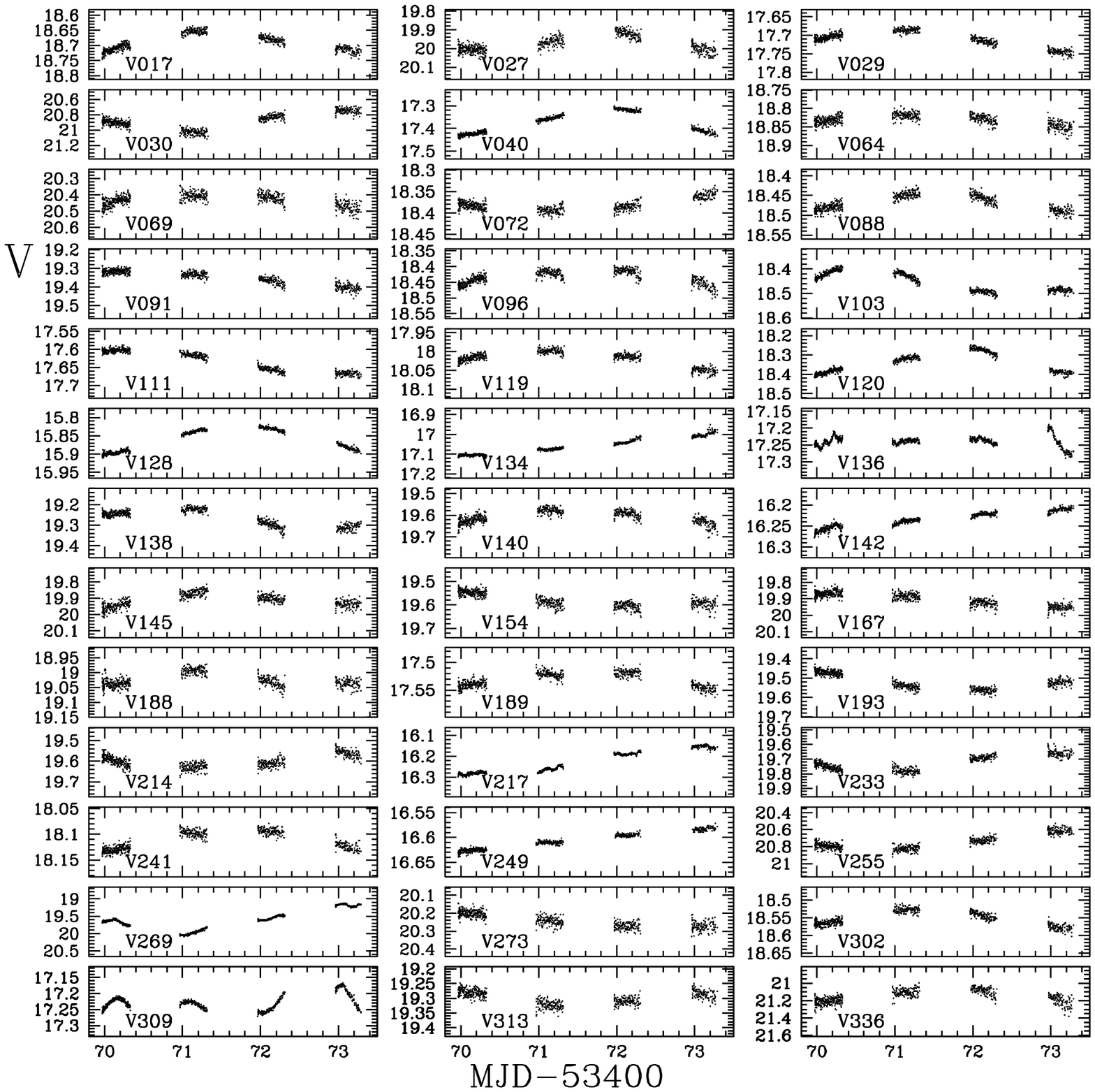}
\caption{Light curves of other variable objects. Most of them are likely
long-period variables.}
\label{pulsother}
\end{figure*}

\subsection{Stars with flares and planetary transits}

Light curves of seven stars with flares are shown in Fig. 15.
Table~1 summaries photometric information on these stars.
The amplitudes of flares in the cases of V007 and V094 are larger
than 2.0~mag, while in the other five cases they are smaller than 1.0~mag.
The fading part of the flare in V294 shows a kind of knee.
Star V033 is periodic, with $P=1.1625$~d (the phased light curve
is shown in Fig.~12).

\begin{table}[h!]
\caption{Photometric information on flares detected in seven stars.}
\smallskip
{\small
\begin{tabular}{ccccl}
\hline
Star & $V$   & $\Delta V$ & Duration & Remarks \\
     & [mag] & [mag] & [h:m] & \\
\hline
V006 & 17.95 & 0.33 & 1:40 & \\
V007 & 22.70 & 2.90 & 1:35 & \\
V033 & 19.32 & 0.20 & 0:55 & periodic, $P=1.1625$~d \\
V082 & 19.47 & 0.15 & 0:55 & \\
V094 & 23.50 & 2.20 & 1:55 & \\
V261 & 19.75 & 0.13 & 0:55 & \\
V294 & 19.92 & 0.62 & 2:50 & \\
\hline
\label{flaretab}
\end{tabular}}
\end{table}

Finally, Fig.~16, presents our own photometry of two planetary
transits: OGLE-TR-111 \citep{pont04} and OGLE-TR-113 \citep{bou04}.
Both of them are caused by hot Jupiters. Details on the photometry
of the transits we give in \cite{pie09}.

\begin{figure*}
\includegraphics[angle=0,width=0.85\textwidth]{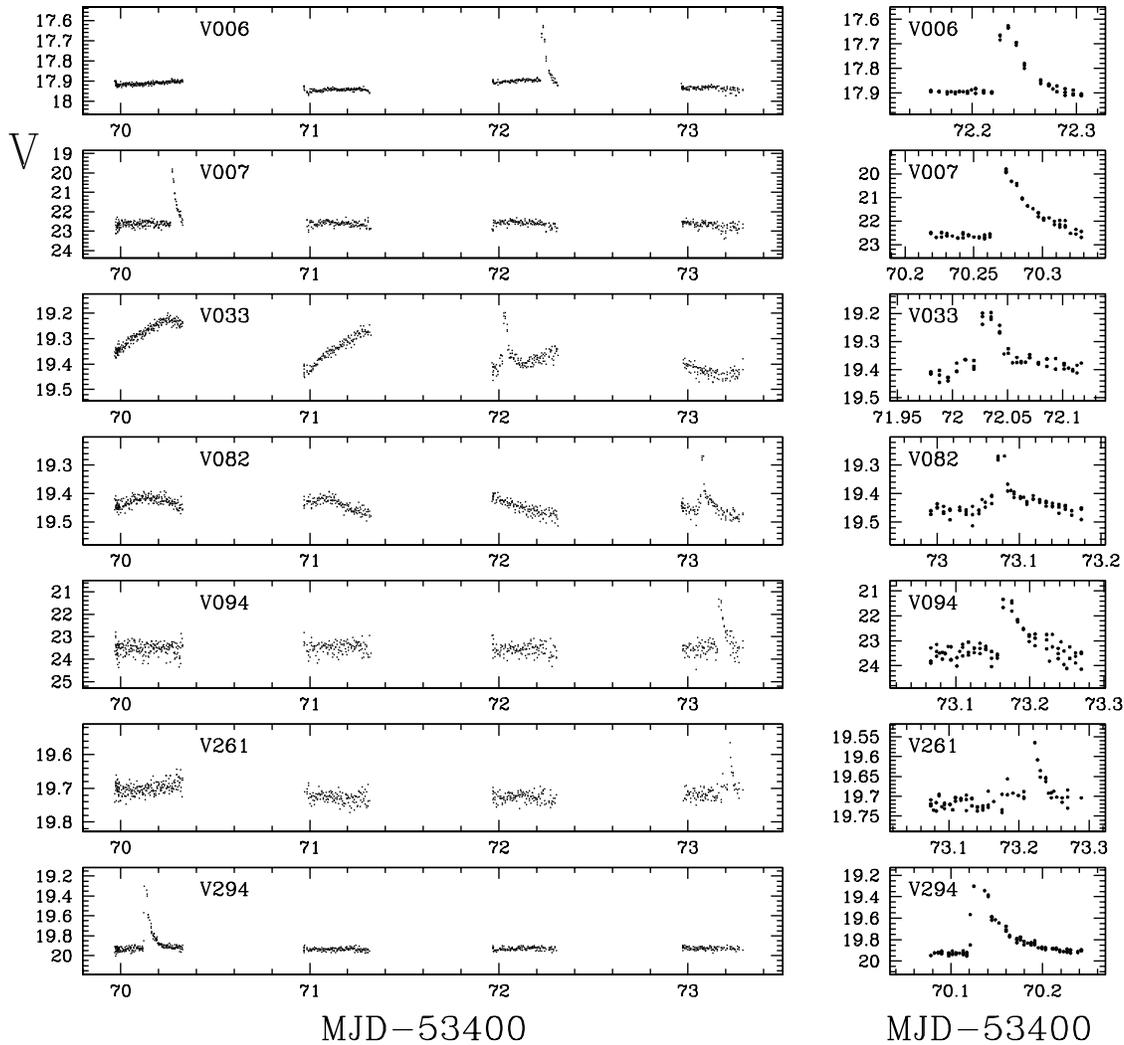}
\caption{Stars with flares. Note the the object V033 is also a periodic (very
likely pulsating) variable with $P=1.1625$~d for which phased light curve
is illustrated in Fig.~12.}
\label{flare}
\end{figure*}

\begin{figure*}
\includegraphics[angle=0,width=0.85\textwidth]{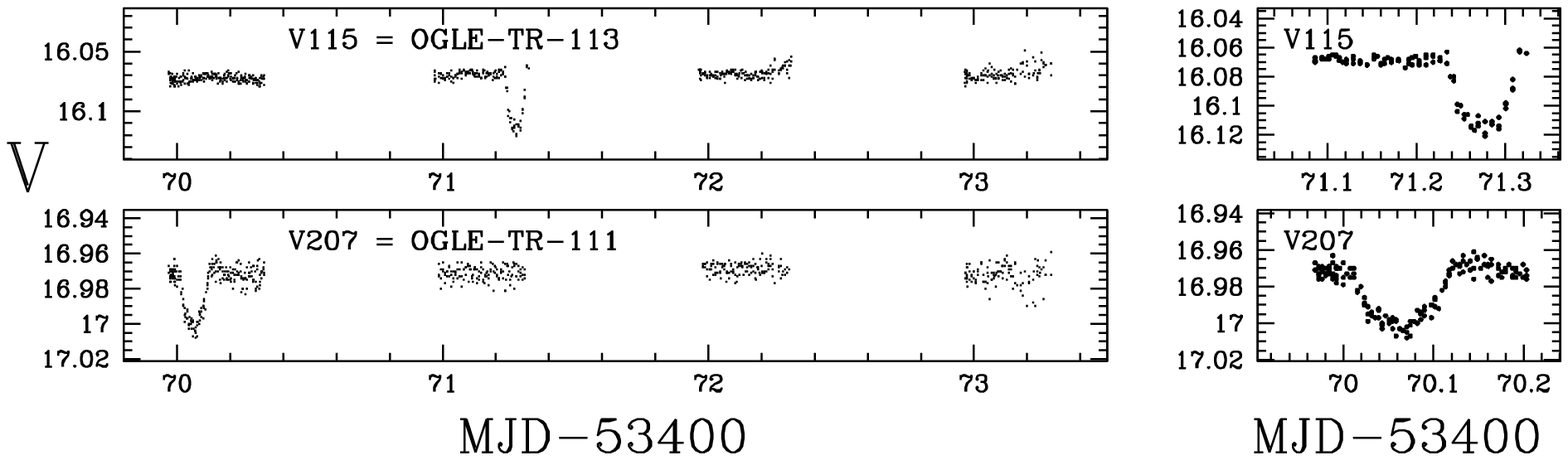}
\caption{Two planetary transits: V115 (OGLE-TR-113) and V207 (OGLE-TR-111).}
\label{planets}
\end{figure*}

\section{Conclusions}

The search for variable objects in the VIMOS field towards the Galactic
plane in Carina resulted in the detection of 348 variables among 50897 stars
down to $V=24.5$~mag. Only five of the objects were previously known
to be variable. These were OGLE transits TR-106, TR-109, TR-110, TR-111
and TR-113. The last two transits were confirmed to be caused by hot Jupiters,
while the first three seem to be binary stars. We note that four other
OGLE objects lie in the same field (TR-105, TR-108, TR-114, TR-198),
but no variability was detected for them during our observations. 
Table~2 gives statistical information on the variables found. All photometric
data, finding charts and a large table with equatorial coordinates, periods,
magnitudes and amplitudes of the variables are available on the Internet at

\begin{center}
ftp://ftp.astro.puc.cl/pub/pietruk/VIMOSvar/
\end{center}

About half of the detected variables are eclipsing/ellipsoidal binaries,
while the other half are pulsating variables of different types,
mostly $\delta$~Scuti stars. Based on the numbers in Table~2 one can
say that about seven stars per 1000 show detectable brightness variations.
On average, three of them are eclipsing/ellipsoidal binaries, three are pulsating
variables (where two are usually $\delta$~Scuti stars). Other variable
objects in the sample show changes on longer time-scales (more than 4~days)
or represent transient events (stars with flares and transits).

\begin{table}[h!]
\caption{Number of variables of different types found in the data.}
\smallskip
{\small
\begin{tabular}{lrr}
\hline
Type of stars                         & Number & Percentage \\
\hline
All stars searched for variability     & 50897 &   100 \% \\
\hline
All known variables in the field       &   352 & 0.692 \% \\
\hline
All variables detected in our survey   &   348 & 0.684 \% \\
\hline
All eclipsing                          &   148 & 0.291 \% \\
Eclipsing with known period            &   121 & \\
Other eclipsing                        &    27 & \\
\hline
Pulsating variables                    &   153 & 0.301 \% \\
$\delta$ Scuti                         &    99 & 0.195 \% \\
Other pulsating                        &    54 & \\
\hline
Other variables                        &    39 & \\
\hline
Stars with flares                      &     7 & 0.014 \% \\
\hline
Planetary transits                     &     2 & 0.004 \% \\
\hline
Variables not detected (OGLE transits) &     4 & \\ 
\hline
\label{varnumtab}
\end{tabular}}
\end{table}

We stress that the VIMOS observations lasted only 4 nights and therefore
we were not able to detect variable objects with longer time scales.
Long periods are often present in red giants, which are bright
and easily detectable.

There has not been any other such deep and of similar time-span and
time-resolution variability survey so far from the ground. However, we can
compare with other two long-term but shallower ground-based surveys recently
published. For example, \cite{wel08} found 494 variables in the sample of 110~372
stars in Lupus, what gives $\sim0.45\%$. The percentage of detected variables
in the ASAS survey \citep{pacz06} is only 0.29\%. There, among 17~000~000
stars 50~099 were found to change their brightness. The percentage of variables
detected in our work, $\sim0.69\%$ of all the stars in the sample, is much higher
than in the two surveys. Also, one has to remember that this is only a lower limit
for the variable detection. A deep Hubble Space Telescope survey dedicated
to find planetary transits was carried out by \cite{sahu06}. Such database would
sample faint variables in the Galactic bulge, useful for comparison with the 
present work.

We decided to search for variables by eye initially just due to the short period
of the VIMOS observations. It is obvious that this method will not be efficient
in case of hundreds of thousands of stars monitored for hundreds of nights.
Our variable star search was later automated, yielding similar results.
The present results, including the large number of newly detected variables
of different types in a limited field, high quality photometry, high accuracy
of the determined periods, show that ground-based wide-field variability surveys
are powerful tool to draw a more detailed picture of our Galaxy.

\begin{acknowledgements}
PP, DM, JMF, GP, MTR, WG, MH are supported by FONDAP Center for Astrophysics
No. 15010003 and BASAL Center for Astrophysics and Associated Technologies PFB06.
MZ and DM also acknowledge support by Proyecto FONDECYT Regular No. 1085278
and 2090213, respectively.
We are grateful to the ESO staff at Paranal Observatory.
\end{acknowledgements}

\end{document}